\documentclass[twoside,leqno,twocolumn]{article}
\usepackage[english]{babel}
\usepackage[T1]{fontenc}
\usepackage[utf8]{inputenc}
\usepackage{microtype}
\usepackage{hyperref}
\usepackage{cleveref}

\usepackage[letterpaper]{geometry}

\usepackage{caption}
\usepackage{subcaption}

\usepackage[algo2e,ruled,vlined,linesnumbered]{algorithm2e}
\Crefname{algocfline}{Line}{Lines}
\usepackage{booktabs}
\usepackage[table]{xcolor}

\usepackage{todonotes}
\usepackage{csquotes}
\usepackage{adjustbox}
\usepackage{placeins}
\usepackage[htt]{hyphenat} 

\newcommand{\algoname}[1]{\textsc{#1}\xspace}

\newcommand{\apq}{\algoname{pq}}
\newcommand{\amatrix}{\algoname{matrix}}
\newcommand{\aset}{\algoname{set}}
\newcommand{\atreap}{\algoname{treap}}
\newcommand{\ahollow}{\algoname{hollow}}
\newcommand{\agabow}{\algoname{ggst}}
\newcommand{\alemon}{\algoname{lemon}}
\newcommand{\aatofigh}{\algoname{atofigh}}
\newcommand{\afelerius}{\algoname{felerius}}
\newcommand{\aspaghetti}{\algoname{spaghetti}}
\newcommand{\ayosupo}{\algoname{yosupo}}

\graphicspath{{./img/}}

\usepackage{ltexpprt}

\begin{document}

\newcommand\relatedversion{}

\title{\Large Efficiently Computing Directed Minimum Spanning Trees\relatedversion}

\author{Maximilian B\"other \thanks{Hasso Plattner Institute.}
\and Otto Kißig \footnotemark[1]
\and Christopher Weyand \thanks{Karlsruhe Institute of Technology.}
}

\date{}

\maketitle







\begin{abstract} \small\baselineskip=9pt Computing a directed minimum spanning tree, called arborescence, is a fundamental algorithmic problem, although not as common as its undirected counterpart.
In 1967, Edmonds discussed an elegant solution.
It was refined to run in $O(\min(n^2, m\log n))$ by Tarjan which is optimal for very dense and very sparse graphs.
Gabow et al.~gave a version of Edmonds' algorithm that runs in $O(n\log n + m)$, thus asymptotically beating the Tarjan variant in the regime between sparse and dense.
Despite the attention the problem received theoretically, there exists, to the best of our knowledge, no empirical evaluation of either of these algorithms.
In fact, the version by Gabow et al.~has never been implemented and, aside from coding competitions, all readily available Tarjan implementations run in $O(n^2)$.
In this paper, we provide the first implementation of the version by Gabow et al.~as well as five variants of Tarjan's version with different underlying data structures.
We evaluate these algorithms and existing solvers on a large set of real-world and random graphs.
\end{abstract}

\section{Introduction}
\label{sec:introduction}

The minimum spanning tree problem is well studied with various applications~\cite{Graham1985,Suk1984} and algorithms~\cite{Jarnik1930, Kruskal1956, Prim1957}.
The directed version, called the minimum spanning arborescence problem, has received much less attention.
For a given root $r$, it aims at finding a directed spanning tree of minimum weight rooted at $r$.
Applications include infection chain modeling~\cite{Jombart2010} and approximating traveling salesperson instances~\cite{Salii2020}.
Different versions and generalizations were studied~\cite{Georgiadis2003, Kamiyama2009, Kamiyama2014}.
Sometimes multiple roots are given or it is required to find the best root.
Historically, the problem was to find a set of non-overlapping trees with maximum total weight, called an optimum branching.
As these versions are linear time equivalent~\cite{Edmonds1967,Mendelson2006}, we focus on the minimum spanning arborescence problem with given root. 

The algorithm to find a minimum spanning arborescence was discovered independently by~Edmonds~\cite{Edmonds1967}, Chu~\cite{Chu1965}, and Bock~\cite{Bock1971}.
Karp~\cite{Karp1971} was the first to give a combinatorial proof of correctness.
Following the literature, we call it Edmonds' algorithm.
The algorithm runs in $O(nm)$ and forms the basis for later, more elaborate versions by Tarjan~\cite{Tarjan1977, Camerini1979} running in $O(\min(n^2,m\log n))$ and Gabow et al.~\cite{Gabow1986} running in $O(n\log n + m)$, which we call the GGST algorithm in the following.
There exist parallel algorithms for different settings of distributed computing~\cite{Lovasz1985,fischer2019}.
They are based on Edmond's Algorithm as well but we will focus solely on the sequential setting.
Both Tarjan's versions and GGST have the same complexity for very sparse and very dense graphs while the GGST version beats Tarjan's by a logarithmic factor for the regime in between.
GGST likely is optimal since the minimum spanning arborescence problem can be reduced to (s,t)-shortest path~\cite{fischer2019} and
comparison based sorting can be reduced to determining the order of contractions performed during Edmonds' algorithm~\cite{Gabow1986}.
However, a time of $O(m\log\log n)$ was obtained in the word RAM model with Tarjan's version~\cite{Mendelson2006}.
Moreover, Tarjan's version was shown to run in $O(n\log^2n+m)$ on Erd\H{o}s-R{\'e}nyi graphs with random weights \cite{Tarjan1977, Erdos1959}.

To the best of our knowledge, no experimental evaluation of these algorithms, or even an implementation of GGST, exist.
The latter is likely due to the rather technical description and the fact that the algorithm is not the main result of the corresponding paper.
On the other hand, there exist some efficient (meaning $O(m\log n)$) implementations of Tarjan's version.
The problem is a niche topic in coding competitions such as the International Collegiate Programming Contest (ICPC).
Unfortunately, they are hard to find because most of them are only documented as submissions in online judge systems.
The only ready-to-use library implementations run in $O(n^2)$.
This paper provides accessible descriptions and implementations as well as a detailed evaluation.
Our code is open source and can be found in our public repository\footnote{\url{https://github.com/chistopher/arbok}}.
The core contributions of this paper include
\begin{itemize}
\item five Tarjan implementations with different underlying data structures, one of which beats existing solvers on most instances,
\item a high level description of the GGST algorithm with several optimizations/simplifications,
\item an efficient implementation of the GGST algorithm,
\item and a detailed experimental evaluation on a large number of real-word and synthetic networks.
\end{itemize}

In Section~\ref{sec:algo} we describe Edmonds' algorithm along with the versions by Tarjan and Gabow et al.
Section~\ref{sec:impl} describes the existing and new implementations and optimization techniques.
The experimental evaluation is presented in Section~\ref{sec:eval}.
We conclude in Section~\ref{sec:conclusion}.

\section{Edmonds' Arborescence Algorithm}
\label{sec:algo}

We discuss Edmonds' algorithm in Section~\ref{sec:edmonds}, Tarjan's version in Section~\ref{sec:tarjan}, and the GGST version in Section~\ref{sec:gabow}.
The latter two yield just the weight of the optimal solution, not the actual edges.
Reconstructing the edge set is discussed in Section~\ref{sec:reconstruction}.

\subsection{Edmonds' Original Version}
\label{sec:edmonds}

Edmonds' algorithm works as follows.
For each vertex $v\neq r$, pick the cheapest incoming edge $\pi(v)$. 
If the set of these $n-1$ edges contains no cycles, it is an arborescence; otherwise, it is possible to show that there is an optimal solution that contains all chosen edges except one for each cycle.
To determine which edge of each cycle to remove, Edmonds' algorithm contracts each cycle.
Note that a vertex is part of at most one cycle.
The weight of all edges going into a cycle $C$ is reduced as follows.
An edge pointing at vertex~$v\in C$ is reduced by the weight of $\pi(v)$, i.e., the weight of the cheapest edge incoming into~$v$.
We then compute a solution on the contracted graph.
The resulting solution has an incoming edge for each cycle $C$ we contracted.
This edge corresponds to an original edge $(u, v)$ with $v\in C$, which we use to replace the cycle edge $\pi(v)$ we picked earlier.

The correctness is based on the following fact. 
Adding a constant $\Delta$ to all incoming edge weights of a vertex changes the weight of each arborescence by $\Delta$, since each solution picks exactly one of those.
This means that the edge cost changes performed during the algorithm preserve the optimal solution.
Moreover, the cycle edges all get a cost of zero. 
Thus, the final cost is the same, no matter which edge is replaced.

\subsection{Tarjan's Version}
\label{sec:tarjan}

Tarjan proposed a version of Edmonds' algorithm that, given the right data structures, runs in $O(m\log n)$ or $O(n^2)$~\cite{Tarjan1977}.
It features two major improvements.
First, the cycle expansion and removal of one edge per cycle is detached from the main algorithm and seen as a postprocessing step.
The algorithm tracks all chosen edges as a superset of the solution,
which can be reconstructed afterwards in linear time.
The second change is to formulate the algorithm sequentially in such a way to avoid rebuilding the graph for each contraction.
The approach goes as follows.
While there is a vertex other than the root that was not processed yet, its cheapest, incoming, edge that is not a self-loop is added to the solution.
If this edge forms a cycle with previously chosen edges, the cost of edges into the cycle is changed as in Edmonds' algorithm and the cycle vertices including their incoming edges are merged into a vertex representing the cycle.
This vertex is then added to the queue of unprocessed vertices.

The algorithm requires data structures to find the cheapest incoming edge, 
recognize cycles of chosen edges, 
and to track contractions.
The latter two can be achieved with disjoint set union (DSU) data structures such as a disjoint set forest~\cite{Galler1964, Tarjan1975}.
To find cycles, a DSU maintains weakly connected components with respect to the chosen edges.
Note that each vertex has at most one incoming chosen edge.
Thus, an edge closes a directed cycle with previously chosen edges, if and only if, it connects two vertices in the same weakly connected component.
A second DSU is used to manage contractions and to map original vertices to contracted vertices.
The endpoints of edges are not updated after each contraction.
Instead, a DSU lookup is required each time the algorithm handles an edge.

The data structure to maintain incoming edge sets must support four operations:
(1) add an element,
(2) extract the minimum element,
(3) change the weight of all elements in the set by a constant, and
(4) merge two sets.
If all operations take at most logarithmic time, the algorithm runs in $O(m\log n)$.
Most mergeable heaps (e.g.,~hollow heaps, treaps, skew heaps) support operations (1), (2), and (4) and can be extended with lazy propagation to allow for operation (3).
Alternatively, if operations 2-4 run in $O(n)$, e.g., when using an adjacency matrix, the algorithm runs in $O(n^2)$, which is better for dense graphs.

\subsection{GGST Version}
\label{sec:gabow}

\begin{figure}
\centering
\includegraphics[width=\columnwidth]{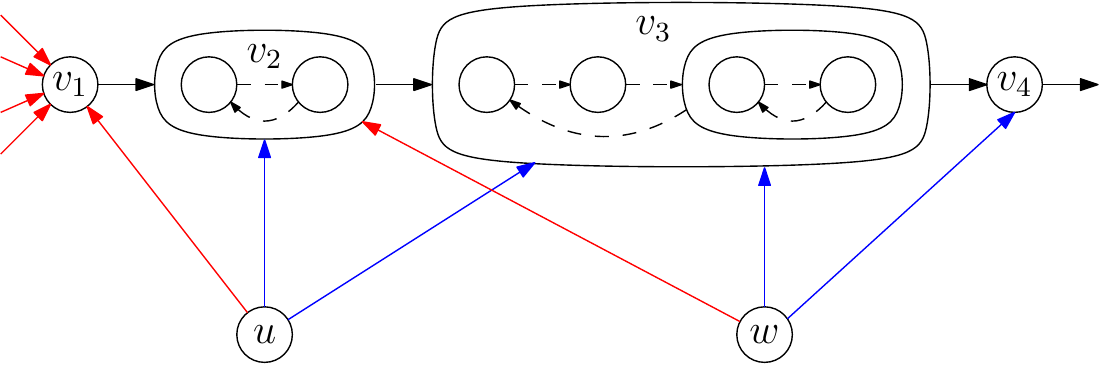}
\caption{Visualization of the growth path with the first four vertices $v_1,v_2,v_3,v_4$. Also shown are the exit lists of two arbitrary vertices $u,w$. Active edges are red and passive edges blue.}
\label{fig:growth-path}
\end{figure}

Gabow et al.~\cite{Gabow1986} further refine the version given by Tarjan to reduce the running time to $\mathcal{O}(n\log n +m)$.
They use the last remaining degree of freedom, namely the order in which vertices are processed.
The authors suggest to always choose the vertex from where the last incoming edge originated thus forming a path of chosen vertices, called the \emph{growth path}.
To avoid special cases for when the path reaches the root, they add dummy edges with cost 0 from the root to all other vertices, making the graph fully connected.
These edges do not affect running time but simplify the description since the algorithm becomes oblivious to the root and the additional edges can be removed in the reconstruction phase.
The improved running time is achieved by exploiting the structure of the path and clever handling of associated edges.
For each vertex, whether on the growth path or not, an \emph{exit list} contains outgoing edges pointing into the growth path.
Exit lists are sorted by the position of their target vertex in the growth path, i.e., the first edge points closest to the head of the path.
The first edge in each exit list is called \emph{active}; all others \emph{passive}.
The active edges are maintained in a data structure we call an \emph{active forest}.
Figure~\ref{fig:growth-path} shows an example.
In the following, we give a concise, yet comprehensive, description of the algorithm that differs from the original discussion in the level of abstraction and simplifies the logic and data structures.

The algorithm starts at an arbitrary vertex and repeatedly picks the cheapest incoming edge of the path head until the path covers the whole graph.
Each iteration, the path is either extended or contracted.
If the origin of the picked edge is not yet on the growth path, then it becomes the new path head.
If it is already on the growth path, then the prefix of the growth path up to this vertex forms a cycle, which is contracted into a single vertex that becomes the new path head.
The process is summarized in Algorithm~\ref{alg:gabow}.
As in Tarjan's version, contractions are tracked with a DSU~\cite{Galler1964, Tarjan1975} which handles the $find(u)$ calls.

\begin{algorithm2e}
initialize growth path with arbitrary vertex\;
insert its incoming edges into exit lists\;
\While{not all vertices on growth path}{
    query min. incoming edge $(u,v)$ of path head from active forest\;
    remember $(u,v)$ for reconstruction\;
    \uIf{find($u$) is not on growth path}{
        insert $u$'s incoming edges into exit lists\;
    }\Else{
        delete prefix of path up to last occurrence of $find(u)$\;
        update incoming edge costs for all vertices on prefix\;
        delete outgoing edges of prefix from exit lists\;
        merge prefix in DSU and Active Forest\;
        limit edges into the cycle to at most 1 per origin\;
    }
    insert $find(u)$ at front of path\;
}
\caption{Minimum arborescence algorithm by Gabow et al.~\cite{Gabow1986}}
\label{alg:gabow} 
\end{algorithm2e}

\subparagraph*{Growth Path Extension.}
When the growth path is extended by a new vertex $u$, all incoming edges of $u$ are introduced to the algorithm and inserted into their respective exit lists.
Consider the insertion of an edge, say $(x,u)$, into $x$'s exit list.
Since $u$ just became the new head of the growth path, the edge will be inserted at the front of the exit list.
It will become active and, if the exit list was not empty, the previously active edge will become passive.
Because $u$ just became part of the growth path, it is not a contracted vertex.
However $x$ may be on the growth path and therefore may be a contracted vertex.
As such, $x$ may have multiple outgoing edges to $u$, originating from different vertices inside $x$.
To deal with this issue (and with multi-edges in the input), one checks if the first edge in the exit list already points to $u$ and if so, only keeps the cheaper one.
This limits the exit list to at most one edge to $u$ maintaining the invariant that an exit list never contains two edges to the same vertex.

\subparagraph*{Growth Path Contraction.}
When a prefix of the path forms a cycle, it is contracted just as in Edmonds' algorithm.
That is, the prefix is removed from the path, incoming edges into the cycle are reduced in cost, edges resulting in self loops are deleted, the cycle vertices are contracted in the DSU as well as in the active forest, and multi-edges are removed.

The cost reduction is done with the DSU, which can be modified to track an offset for each vertex~\cite{Gabow1986}.
Whenever the current cost of an edge is needed, a DSU lookup analogous to a $find$ is made to get the offset of the target vertex.

Self loops are outgoing edges from the cycle, so by deleting all edges in exit lists of cycle vertices, self loops are avoided.
This also deletes edges pointing further down the path but these are irrelevant to the algorithm.
They can only become incoming edges of the head if, in the future, the path is contracted up to their target and in this case they would be self loops.

Edges that became multi-edges by the contraction are consolidated.
For each vertex with more than one edge into the cycle, the prefix of their exit list that points into the cycle is deleted except for the cheapest of those edges.
If a vertex has more than one edge into the cycle, at least one of them is passive.
Thus, such vertices can be found by maintaining for each vertex on the growth path a list of incoming passive edges, called a \emph{passive list}.

\subparagraph*{Active Forest.}
The active forest maintains all currently active edges and must be updated accordingly.
It stores for each vertex the outgoing active edge and a set of incoming active edges.
We associate an active edge with the vertex it originates from.
The active forest is able to 
INSERT an active edge for a vertex that does not yet have one in $O(1)$,
REPLACE the active edge of a vertex by another one that points closer to the growth path head or points to the same vertex but has less weight in $O(1)$,
DELETE the active edge of a vertex in $O(\log n)$,
MERGE the sets of incoming active edges for the first two vertices of the growth path in constant time,
and QUERY the minimum incoming active edge of the path head in $O(\log n)$ amortized time.

It is implemented as follows.
Each vertex stores its incoming active edges in a Fibonacci heap~\cite{Fredman1987}, which enables the operations
INSERT, DELETE, MERGE, and QUERY by just mapping them to the corresponding Fibonacci heap operations.
The REPLACE operation could be implemented as a DELETE followed by an INSERT.
Unfortunately, this results in a running time of $O(\log n)$.
Instead, Gabow et al.~\cite{Gabow1986} suggest to reuse the internal heap node representing the old edge. 
The node is \emph{moved} from the heap the old edge is in to the heap where the new edge should be and receives the new edge as key.
This \emph{move} takes $O(1)$ time and is the crucial point where the logarithmic factor over Tarjan's version is saved.
The move operation is possible by restricting QUERY, REPLACE, and MERGE to the structure of the growth path.
In general, no mergable heap data structure is known that lifts these restrictions and still supports something like a move~\cite{Mendelson2006}.

However, the \emph{move} has two major problems for which we need to understand some internals about Fibonacci heaps.
A Fibonacci heap is a forest whose roots are kept in a list called the \emph{root list} of the heap.
Each tree maintains the heap property, i.e., the key of a child node is higher or equal to the key of its parent.
The key in our case is the weight of the corresponding active edge.
Also, a Fibonacci heap usually maintains the minimum key of nodes in the root list to allow queries in constant time.
The first problem of the move operation is that the cached minimum of a root list cannot be updated in constant time if the current minimum is moved out of that list.
Therefore, we do not maintain the minimum. 
Instead, the QUERY operation rebuilds the root list, which is a common operation for Fibonacci heaps usually done upon extraction of the minimum, resulting in an amortized $O(\log n)$ running time.
The second problem is that moving an internal heap node actually moves the whole subtree rooted at this node.
Descendants of the node are displaced into the wrong heap and, moreover, changing the key of the moved node can violate the heap property.
To fix the displacement, every time a Fibonacci heap operation would put a node into the root list they are returned to the root list of the heap the node actually belongs to, which we call the \emph{home heap} of that node.
That is, the \emph{home heap} of a heap node is the heap of the target vertex of the corresponding active edge.
Finding the home heap requires a DSU lookup because the target vertex might be contained in a contracted vertex.
Gabow et al.~\cite{Gabow1986} prove the following three invariants to address the violated heap property and the correctness of the home heap fix.
(1)~The root of any tree is always in its home heap.
(2)~The heaps maintain an additional heap property w.r.t.~their home heaps ordered by the position in the growth path.
That is, the home heap of a parent node is at least as close to the growth path head as the home heaps of its children.
(3)~The original heap property is never violated between two nodes that are in their home heap.
Only displaced nodes can temporarily violate the heap property.

\subparagraph*{Time Complexity.}
The growth path is extended at most $n$ times. 
Since contracting a cycle of length $l$ reduces the total number of vertices in the graph by $l-1$, 
there are at most $n-1$ contractions and the summed length of all contracted cycles is less than $2n$.
Thus, QUERY, DELETE, and MERGE are called $O(n)$ times on the active forest.
Furthermore, the following operations happen at most once per edge and can all be done in constant time.
Insertion into an exit list, the active forest, or passive list, REPLACE in the active forest, deletion from an exit list, and deletion from a passive list.
The DSU imposes no additional overhead since, if there are at least $n\log n$ calls to $find$, each individual one takes amortized constant time~\cite{Tarjan1975}.
In total this yields a running time of $O(n\log n + m)$.

\subsection{Arborescence Reconstruction}
\label{sec:reconstruction}

\begin{figure}
\centering
\includegraphics[width=\columnwidth]{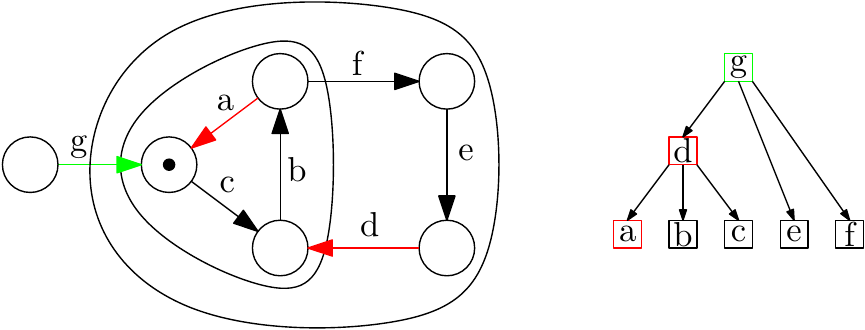}
\caption{Example graph (left) and corresponding reconstruction forest (right). Assume Gabow starts at the vertex marked with a dot and the edges are labeled alphabetically in the order they were added to the growth path. The first step of the reconstruction process is indicated by colors.}
\label{fig:recon}
\end{figure}

Although Tarjan~\cite{Tarjan1977} proposed to split off the reconstruction phase from main algorithm,
the reconstruction method given in the paper is incorrect.
A note by Camerini et al.\ outlines a working method~\cite{Camerini1979}.
Consider a new graph called the \emph{reconstruction forest} where the nodes are the edges that are picked by the arborescence algorithm.
In the forest, an edge that was picked as an incoming edge to a contracted vertex has directed arcs to the edges that constitute the top-level cycle of the contracted vertex.
A leaf in the reconstruction forest corresponds to the first picked incoming edge of a vertex of the original graph.
Figure~\ref{fig:recon} shows an example.
The reconstruction process repeatedly selects a root of the forest.
The corresponding edge becomes part of the final solution.
The target vertex in the original graph of the selected edge has an associated leaf in the reconstruction forest.
The process deletes the path from this leaf to the selected root from the forest, then proceeds with the next root.

A very concise implementation is possible by noting that the order in which the main algorithm picks edges is a reverse topological order of the reconstruction forest.
Thus, all roots are found by iterating over the picked edges in reverse and skipping already deleted ones.
Further required information are the leaf of each original vertex and the parent for each node in the reconstruction forest.
The former is computed by iterating over the picked edges to find the first occurrence of each target.
The latter must be saved by the main algorithm each time it contracts a cycle.

\section{Implementation}
\label{sec:impl}

This section introduces different solvers for the minimum arborescence problem, highlights their key points as well as optimizations and deviations from the abstract description in Section~\ref{sec:algo}.
We compare 11 solvers; our five versions of Tarjan's approach using different data structures, five external Tarjan-based solvers, and our Gabow implementation (see Table~\ref{tab:impls}).
External solvers fall into two categories, coding competition code and library solvers.
We refer to \Cref{appendix} for more discussions of the external solvers, e.g., integration issues and coding errors we had to fix.

\begin{table*}
  \definecolor{tablegray}{gray}{0.9}
  \definecolor{othergray}{gray}{0.6}
  \rowcolors{1}{white}{tablegray}
  \centering
  \caption{Overview of arborescence algorithms. Tarjan+Path means they implement Tarjan's version but adjust the order in which vertices are processed to form a path as in Gabow's version.}
  \begin{tabular}{l c c c c}
    \toprule
    Solver & Author/Source & Variant & Data Structure & Runtime \\
    \midrule
    \afelerius & David Stangl & Tarjan & Skew Heap~\cite{Sleator1986} & $O(m\log n)$ \\
    \aspaghetti & Takanori Maehara & Tarjan+Path & Skew Heap~\cite{Sleator1986} & $O(m\log n)$ \\
    \ayosupo & Kohei Morita & Tarjan+Path & Pairing Heap~\cite{Fredman1986} & $O(m\log n)$ \\
    \alemon & LEMON 1.3.1 & Tarjan+Path & Adjacency List & $O(n^2)$ \\
    \aatofigh & Ali Tofigh & Tarjan & Adjacency List & $O(n^2)$ \\
    \amatrix & this paper & Tarjan & Adjacency Matrix & $O(n^2)$ \\
    \atreap &  this paper & Tarjan & Treap~\cite{Seidel1996} & $O(m\log n)$ \\
    \ahollow &  this paper & Tarjan & Hollow Heap~\cite{Hansen2017} & $O(m\log n)$ \\
    \aset &  this paper & Tarjan & Red Black Tree~\cite{Guibas1978} & $O(m\log^2 n)$ \\
    \apq &  this paper & Tarjan & Binary Heap~\cite{Cormen2009} & $O(m\log^2 n)$ \\
    \agabow & this paper & GGST & Fibonacci Heap~\cite{Fredman1987} & $O(n\log n + m)$ \\
    \bottomrule
  \end{tabular}
  \label{tab:impls}
\end{table*}

\subparagraph*{Competition Codes.}
Coding competitions occasionally feature arborescence tasks which require an efficient implementation for sparse graphs.
These implementations are often not as maintainable or usable as library solvers, but written with a high focus on performance.
The online judge platform Library Checker\footnote{\url{https://judge.yosupo.jp/problem/directedmst}} contains a test set for the minimum arborescence problem.
We include the jury solution by the maintainer Kohei Morita as well as the fastest submission by David Stangl.
We denote them by \ayosupo\footnote{\url{https://codeforces.com/profile/yosupo}} and \afelerius\footnote{\url{https://codeforces.com/profile/Felerius}} according to their pseudonyms on popular contest websites.
Furthermore, there is a competition-style implementation by Takanori Maehara\footnote{\url{https://github.com/spaghetti-source/algorithm}}, which we denote by \aspaghetti.
With around 130 lines of code, it is the most concise implementation.
However, it lacks the reconstruction phase and proper memory management.

\subparagraph*{Library Solvers.}
The two library implementations we consider are \alemon and \aatofigh, both running in $O(n^2)$.
The former is part of the LEMON library for graph algorithms\footnote{\url{https://lemon.cs.elte.hu/trac/lemon}}.
We use the latest release 1.3.1 from 2014.
They save incoming edges in arrays.
The merge is done by iterating over all incoming edge lists of cycle vertices while collecting the cheapest edge into the cycle for each origin.
They reuse the same collecting array each time and clear the used entries afterwards, such that the merge is not $O(n)$ but linear in the number of merged edges. Thus, the solver is faster the less edges are involved in each contraction.
The second library implementation, \aatofigh, was written by Ali Tofigh and Erik Sjölund\footnote{\url{https://github.com/atofigh/edmonds-alg}} using the Boost Graph Library~\cite{Schaeling2011}.
They also represent incoming edge sets in dynamically growing arrays.
However, the arrays are sorted by origin vertex and the merge is done with the linear time merge routine usually known from merge sort.
It was modified to remove multi-edges by only keeping the cheapest one for each origin.
The same performance considerations apply.
Note that there exist sparse networks where these merge strategies yield quadratic running time.

\subparagraph*{Our Tarjan-based Solvers.}
Our Tarjan code shares the logic for the algorithm and reconstruction and differs only in the data structure to manage the sets of incoming edges (see Section~\ref{sec:tarjan}).
The \amatrix solver maintains an adjacency matrix and performs the operations in linear time.
The \ahollow and \atreap solvers use our implementations of Hollow heaps~\cite{Hansen2017} and Treaps~\cite{Seidel1996}, respectively, which both support lazy propagation to update weights.
The hollow heap is not required to implement the usual decrease key operation, as it is not required by the algorithm, which allows for implementing the merge operation efficiently, by simplifying some bookkeeping tasks.
The \aset and \apq variants use the \texttt{std::set} and \texttt{std::priority\_queue} data structures from the C++ standard template library.
They are typically implemented as a red-black tree~\cite{Guibas1978} and binary heap~\cite{Cormen2009}, respectively.
Since the set and priority queue interface do not support a fast merge operation, we use the well known \emph{smaller into larger} technique.
That is, for a merge we iterate over the smaller of the two sets and add the elements individually to the larger set.
An element switches sets at most $O(\log n)$ times, each time into a set that is at least twice as large, and a switch takes $O(\log n)$.
This sums up to $O(m\log^2n)$ for all merges combined.
Since the elements are moved individually, weight updates do not need lazy propagation but are handled by an offset for each set that is applied when an element enters or leaves the set.

\subparagraph*{Our GGST Solver.}
The solver features three optimizations compared to the description in Section~\ref{sec:gabow}.
First, no dummy edges are inserted. 
Instead a new path is started each time the root is reached.
Second, we replace linked lists by dynamic arrays where possible. 
Exit lists, passive lists, and the growth path are only modified at the front, so an array can be used by saving them in reverse.
Actually, the usage of passive lists as previously described requires arbitrary deletions and thus cross references for each edge to the position in the list.
Our third optimization is to remove the need for cross references by simplifying the deletion patterns.
The only time the algorithms deletes edges is during the contraction of a cycle\footnote{The original description by Gabow et al.~has more deletions. We simplified the algorithm in this regard.}.
Outgoing edges are deleted by clearing complete exit lists and mirroring the deletions across passive lists. 
Incoming multi-edges are deleted by clearing complete passive lists and mirroring the deletions across exit lists.
We modify the two steps to make synchronization between exit and passive lists easier and restrict modifications to the front of the lists.

When outgoing edges of a cycle are deleted, some of these edges are self loops and some point further down the growth path.
Instead of mirroring the clearing of the exit lists by deleting corresponding entries from passive lists, we suggest to entirely skip the removal from the passive lists.
This, of course, keeps invalid entries in the passive lists.
However, a passive list is only read during a contraction to identify multi-edges into the cycle.
At this time, the invalid entries point into a prefix of the path but at the time of deletion pointed down the path.
Thus, they became self-loops which can be identified and skipped.
Since a passive list is cleared after identification of self loops, each invalid entry is seen only once.

We propose to implement the consolidation of multi-edges as follows. 
For each passive edge into the cycle, compare the first two edges in the exit list of the origin of the passive edge and delete the more expensive one.
This \enquote{delete one of the first two edges} operation is done for each origin as often as this origin has passive edges into the cycle.
Since this origin's exit list starts with an active edge pointing into the cycle, followed by all the passive edges into the cycle, the cheapest edge of this prefix will remain at the front of the exit list.
Gabow et al.~propose a similar strategy but delete either the first edge or the currently inspected passive edge (instead of the second in the exit list), which requires for each passive edge a way to obtain its handle in the exit list.

\section{Experiments}
\label{sec:eval}

In this section we evaluate the solvers listed in Table~\ref{tab:impls}.
The solvers, data preparation scripts, plotting code, execution logs, and the raw timing data are available in our public repository.

\subparagraph*{Setup.}
The experiments were performed on a server with two 8-Core Intel Xeon\texttrademark\ Gold 6144 CPUs and 192\,GB DDR4 memory on the openSUSE Leap 15.3 operating system.
The implementations are written in C++ and adjusted to fit a common interface.
The code was compiled with gcc version 10.3.0.
Each run had a timeout of 30 minutes.
We used a total of 656 networks from the following sources. The number of networks is in parenthesis.
\begin{itemize}
\item \textbf{konect} (319). All directed networks smaller than 5GB from the KONECT project\footnote{\url{http://konect.cc/}}.
\item \textbf{networkrepository} (75). A selection of sparse networks from the Network Repository project\footnote{\url{https://networkrepository.com}}. The project contains mostly undirected networks and does not label directed ones as such. We downloaded all networks (around 3000) and kept the ones that are labeled as directed in their respective file format.
\item \textbf{girgs} (200). This data set contains Geometric Inhomogeneous Random Graphs (GIRGs), a generative network model closely related to hyperbolic random graphs~\cite{Bringmann2019,Krioukov2010}.
We used the efficient generator by Bl\"asius et al.~\cite{Blaesius2019} with default parameters except for \(n\), \(deg\), and seeds.
We set \(n=10^4\) and average degrees from 50 to 2000 in steps of 100 with 10 networks per configuration. Edges are directed randomly.
\item \textbf{antilemon} (5). A sparse family of networks crafted to be difficult for arboresence solvers. They require at least $n/2$ contractions with at least $n/2$ edges pointing into each contracted cycle. We generated networks with $n=10^i$ for $i\in[2,6]$. 
\item \textbf{fastestspeedrun} (47). Test cases of a programming task from the ICPC Northwestern Europe Regional Contest 2018\footnote{\url{https://2018.nwerc.eu/}}. They have up to 2500 vertices and are fully connected.
\item \textbf{yosupo} (10). Test cases for the Directed MST problem on the Library Checker website. The networks are Erd\H{o}s-R\'enyi graphs~\cite{Erdos1959} with a random spanning tree from the root vertex as subgraph. Weights are sampled uniformly at random.
\end{itemize}
For networks without weights, we sample integer weights uniformly at random.
If an instance has no specified root, we restrict us to the largest connected component, and add a root vertex that connects to all original vertices with edges of weight infinity.

\begin{figure}
\centering
\includegraphics[width=\columnwidth]{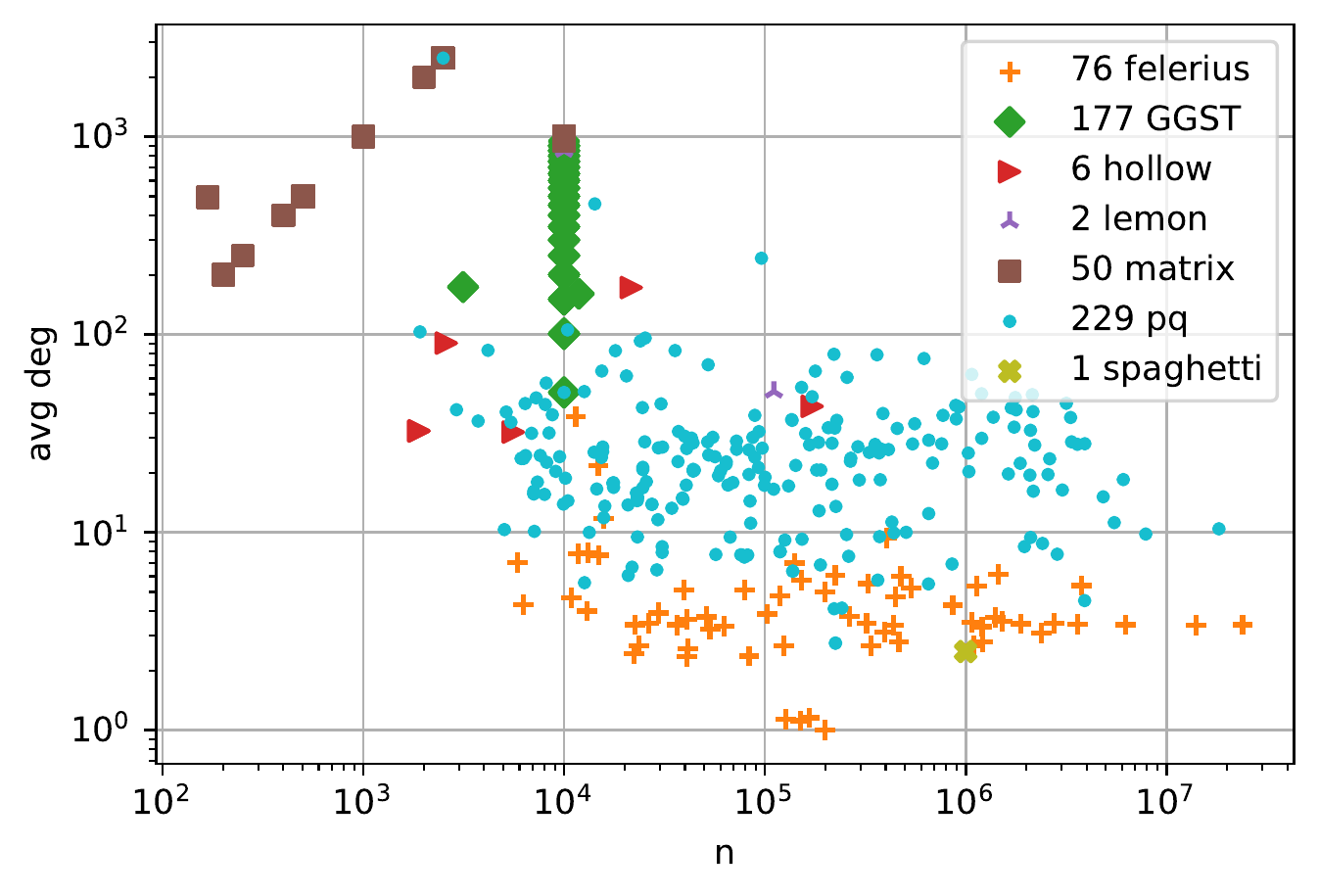}
\caption{For each instance the untied fastest solver if any. The legend includes the number of wins per solver.}
\label{fig:fastest}
\end{figure}

\subparagraph*{General Performance.}
\Cref{fig:fastest} shows all instances with an untied fasted solver, i.e., a solver that is strictly faster than all others.
The major reason for ties is that two or more solvers are faster than 1\,ms which is the precision of our measurements.
Over all, 115 instances are tied,
85 instances have at least two algorithms that solve the instance faster than 1\,ms,
73 instances are solved in under 1\,ms by at least six solvers, and 46 instances are solved in under 1\,ms by all solvers.

On the untied instances, the \apq solver dominates with 229 wins, followed by \agabow with 177, then \afelerius with 76, and \amatrix with 50.
Combined, these four solvers win more than 98\,\% of the untied instances.
Moreover, there is a clear trend regarding the type of instance each solver is good at mirroring the theoretical complexities of the algorithms quite closely.
The matrix-based Tarjan solver, \amatrix, is best for dense graphs, the heap-based Tarjan solvers, \apq and \afelerius, are optimal for sparse graphs, and the GGST algorithm wins in between.
For the sparse real-world instances, there is a clear cut between the \apq and \afelerius solvers.
The \afelerius solver was specifically tuned to be fast on the yosupo instances which have barely more edges than vertices and thus wins on instances with average degree below 10.
Furthermore, all but three of the 177 \agabow wins are on GIRGs.
We explicitly generated the GIRGs to fill the gap between the sparse real-world networks and the fully connected fastestspeedrun instances.
The most surprising result, however, is that the \apq solver using a binary heap performs exceptionally well although it should scale worse in the number of edges than the competitors by at least a logarithmic factor due to the missing merge operation.
We identify three possible reasons for this behavior.
First, a binary heap implementation is very efficiently while the more complex logic of \agabow and less cache efficient data structures of \afelerius cause significant overhead.
Second, realistic data is easy in the sense that the contractions, which are the theoretical bottleneck of the \apq solvers, occur not as often or involve less edges and vertices.
Finally, realistic networks are sparse and thus $O(n\log n)$ becomes indistinguishable from $O(m\log n)$, which is the remaining complexity of the binary heap implementation when ignoring the cost for contractions.
Therefore, on sparse networks with few contractions, the three solvers \agabow, \afelerius, and \apq all have a complexity of roughly $O(n\log n)$ and it comes down to implementation details like memory layout, cache efficiency, and the level of code optimization.
For the same reason \amatrix beats \agabow on very dense instances where both solvers have a complexity of $O(n^2)$.

\begin{figure}
\includegraphics[width=\columnwidth]{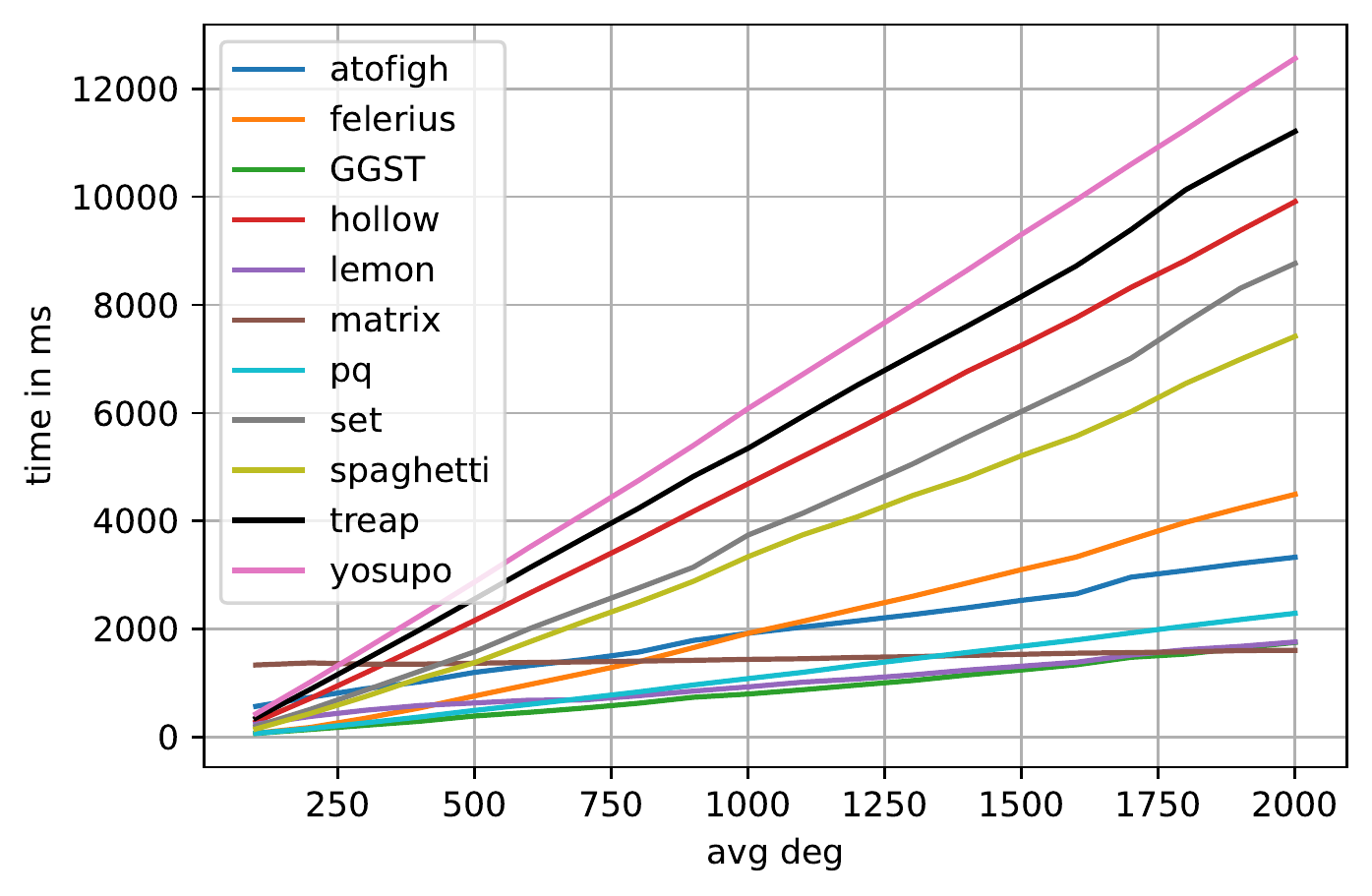}
\caption{The run time of the solvers on GIRGs with $10^4$ vertices over growing density. Each data point is averaged over 10 GIRGs with the same density.}
\label{fig:time-density}
\end{figure}

\subparagraph*{Scaling Analysis.}
To examine the effect of density on solver performance we use the girgs data set.
The GIRG model produces realistic networks regarding degree distribution, clustering, and distances that resemble the real-world networks from the networkrepository and konect data sets.
\Cref{fig:time-density} shows the results.
As expected, the \amatrix solver is not affected by the number of edges.
It starts out as the slowest solver but beats all the others by the time the degree reaches 2000.
All other solvers exhibit an approximately linear scaling in the number of edges which emphasizes again that logarithmic factors are hardly noticeable for reasonably sized inputs.
Most notably, this includes the $O(n^2)$ \aatofigh and \alemon solvers.
These solvers heavily depend on the fact that the instance structure is easy and needs few contractions involving few edges.
The \alemon solver is the second fastest solver only slightly outperformed by \agabow indicating that GIRGs are even easier to solve than the real-world networks from the other data sets.
The reason for this could be the randomized edge direction for the GIRGs.
Another interesting fact is that the five solvers that scale the worst with growing density are \ayosupo, \atreap, \ahollow, \aspaghetti, and \aset.
These five have in common that they use pointer-based heap data structures to manage the edges. 
The other solvers use indices into a preallocated pool (\afelerius), don't have a heap element for every edge (\agabow), or don't use a heap to manage edges (\alemon, \aatofigh).

\begin{figure*}
\centering
\includegraphics[width=\textwidth]{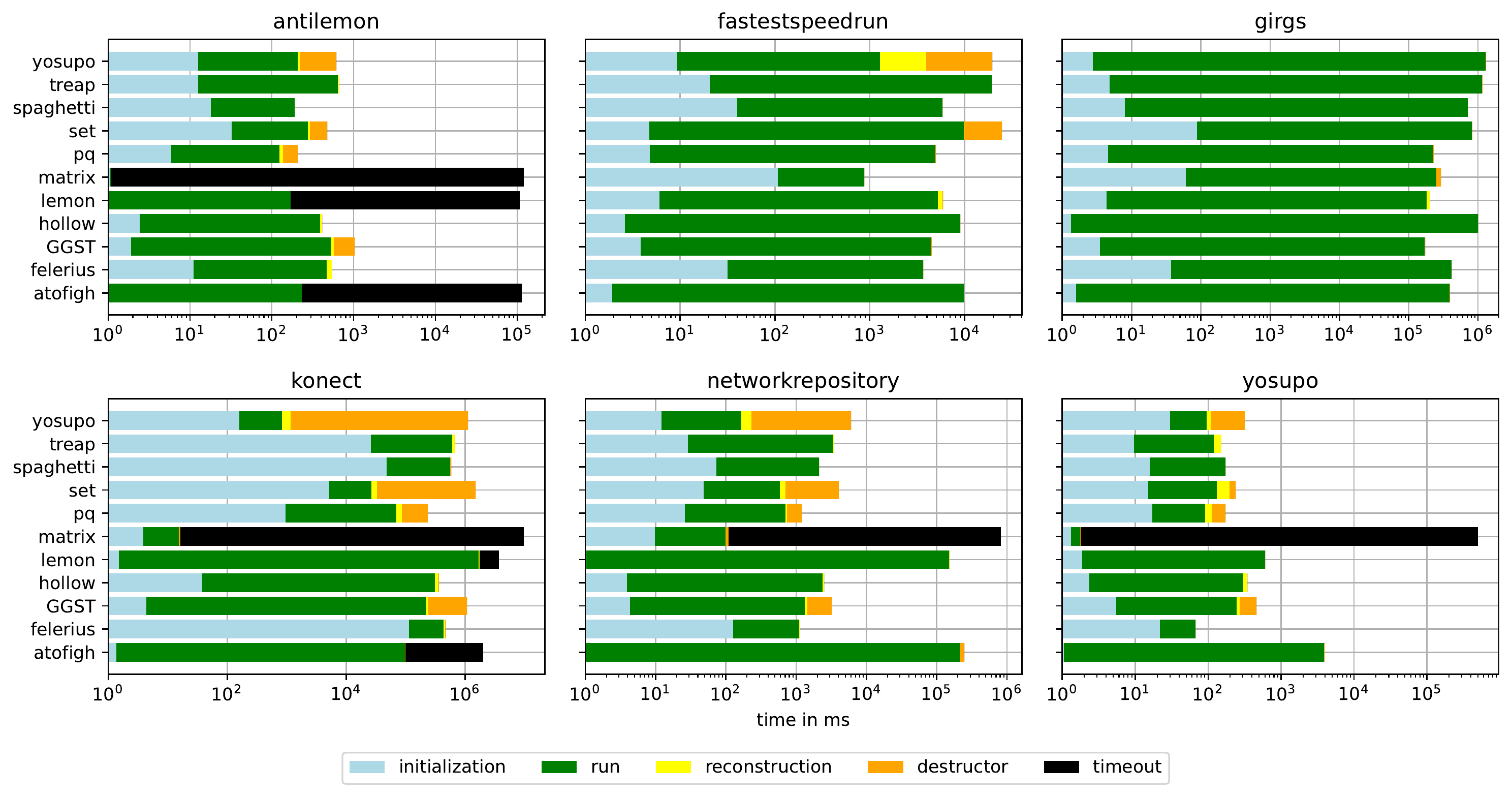}
\caption{For each data set the summed run time over the contained instances per algorithm. The bars are divided into colored segments to show the fraction of time spend on each subroutine. For timeouts, all 30 minutes are counted as timeout no matter what was done in these 30 minutes. Note that the colored segments inside each bar are completely detached from the logarithmic x-axis. 
}
\label{fig:time-bar}
\end{figure*}

\subparagraph*{Time Per Operation.} 
\Cref{fig:time-bar} shows the run times of the solvers divided into initialization, execution, reconstruction, and destructor subroutines as well as timeouts.
The \aatofigh solver crashed by exceeding the available memory on the largest antilemon graph and 8 road networks from the konect data set, which are originally from the 9th DIMACS Implementation Challenge on shortest paths.
These crashes are treated as timeout.
The \amatrix solver is only executed on graphs with less than $10^5$ vertices, and treated as timeout otherwise.
The \alemon solver timed out on three DIMACS graphs and the largest antilemon graph.

On the real-world networks from the konect and networkrepository data sets, the quadratic solvers perform much worse than the other algorithms.
Of course \amatrix cannot handle large graphs but also \aatofigh and \alemon occasionally encounter a difficult instance that overshadows their good performance on the many easy instances since we only consider the summed run time here.
The \alemon solver performs well on the girg data set and the \amatrix solver dominates the fully connected fastestspeedrun instances.
Otherwise the quadratic solvers are never among the fastest.
In particular, these three solvers are more than two orders of magnitude slower than the \afelerius solver on the networkrepository graphs where the \afelerius solver is the fastest on all instances.
The antilemon instances were crafted as worst case instances for \alemon and \aatofigh which is clearly visible in the results.
On this data set, the \apq solver outperforms the others.
Unsurprisingly, the \afelerius solver performs best on the yosupo data which it was optimized for.

Before evaluating the run times of individual subroutines, we note that \alemon and \aatofigh perform most of the initialization and reconstruction operations in the main phase of the algorithm while \aspaghetti has neither reconstruction nor memory management.
With that in mind, our experiments show that the reconstruction phase takes only a fraction of the run time independent of solver or data set.
Furthermore, the initialization, which includes allocating memory and building internal data structures, takes a considerable amount of time for all algorithms.
The high initialization time can be explained by the fact that just inserting the edges into the heap data structures takes $O(m\log n)$ and as such is one of the theoretical bottlenecks of most implementations.
There exist linear time constructions for some of the data structures (e.g. treaps, binary heaps, skew heaps) but for consistency across solvers, we build them by repeated insertions.
Finally, the high destructor time of the \ayosupo solver is due to their use of \texttt{std::shared\_ptr} instead of manual memory management.

\section{Conclusion}
\label{sec:conclusion}
In this paper we discussed the Tarjan and GGST versions of Edmonds' algorithm for the minimum spanning arboresence problem.
We outlined existing solvers, provide our own implementations, and compare their practical performance.
Our implementation of the GGST algorithm is the first public implementation and our description simplifies the original one in several aspects.
Our experiments show that the compared solvers perform well on real-world data while scaling experiments suggest that realistic networks are substantially easier than worst-case instances.
Even solvers with an $O(n^2)$ worst-case complexity often perform almost linear in the number of edges.
However, they are not as consistent.
They time out when the instance contains difficult structures, which occasionally happens even on real-world networks.
Furthermore, we find that differences in complexity by logarithmic factors are mostly irrelevant in practice.
Our $O(m\log^2n)$ Tarjan implementation using a binary heap beats the other solvers on most real-world networks although our $O(n\log n+m)$ GGST implementation is two logarithmic factors faster asymptotically. 
This, again, emphasizes that real-world instances often do not force the worst case of an algorithm and complex logic and data structures can produce significant overhead.
For future work, it would be interesting to examine what makes realistic instances easy and possibly show a better running time on a random model like hyperbolic random graphs similar to the result for Erd\H{o}s-R{\'e}nyi graphs.

\FloatBarrier
\bibliography{ms}

\clearpage
\appendix

\section{Appendix}\label{appendix}

In this section, we briefly describe some further details on the external solvers.

\subparagraph*{Integration Issues and Bug Fixes.}

The \alemon solver does not compile with C++20 upwards because it uses allocator methods that were deprecated in C++17 and removed in C++20.
It performs reconstruction during the main algorithm.
In Figure~\ref{fig:time-bar}, its reconstruction time is the time to obtain the solution from their internal data structures.

The \aatofigh solver contains a programming error in a radix sort subroutine where a right shift equal to the size of the left hand operand type (\texttt{int} in our template instantiation) is performed.
The C++ standard\footnote{The standard must be purchased but a working draft is available at \url{http://www.open-std.org}}
states in Section 7.6.7 concerning shift operators ``The behavior is undefined if the right operand is negative, or greater than or equal to the width of the promoted left operand''~\cite{cpp20}.
Most compilers give a warning (if enabled) and default to 0, which actually works with the given implementation.
Nevertheless, we fixed this error by changing $\leq$ to $<$ in the loop that iterates over the radix.
The atofigh solver performs reconstruction during the main algorithm.
In Figure~\ref{fig:time-bar}, its reconstruction time is the time to obtain the solution from their internal data structures.

The \ayosupo solver uses \texttt{std::shared\_ptr} for memory management.
On very large instances this crashes due to a stack overflow caused by deep recursion in the destructor.
On Linux machines, one can increase the stack limit to circumvent this problem, which is what we do in our experiments.

The \aspaghetti solver does not free allocated memory which gives it an advantage over the other solvers.
We decided to keep the leak since a proper cleanup would require considerable changes to their code.
The \aspaghetti solver is the only solver that does not support reconstruction.

\subparagraph*{Alternative Reconstruction by Stangl.}
The \afelerius solver features an alternative method for reconstruction more closely related to the original idea of Edmonds' algorithm. 
Recall that Edmonds' algorithm contracts each cycle $C$ and when picking an incoming edge into the contracted vertex, it replaces one of the cycle edges.
That is, the edge into the contracted vertex corresponds to an original edge $(u,v)$ and replaces the cycle edge incoming to $v$.
The difficulty when adapting this to Tarjan's version is that endpoint indices of edges are not explicitly updated after each contraction.
Thus, one has to deal with the possibility that the cycle vertices are contracted vertices representing previous cycles.
In this case, $v$ might be contained in a cycle vertex $v'\in C$ rather than being part of the cycle itself.
This is, e.g.,~the case in Figure~\ref{fig:recon} where the edge $g$ replaces the edge $d$.
Stangl tackles this challenge as follows.
Since Tarjan maintains an incoming edge for each vertex during the main algorithm, the reconstruction phase processes the cycles from last to first and performs the necessary replacements.
When a cycle is processed, the edge $(u,v)$ that was picked as incoming for this cycle can be found as the incoming edge to the vertex representing the cycle.
To find the cycle edge it should replace, a persistent DSU is used to query the cycle vertex $v'$ containing $v$ at the time just before the cycle was contracted.
To make the DSU persistent, Stangl drops path compression~\cite{Tarjan1975} from the data structure which means each \emph{find} call takes $O(\log n)$.
However, the main algorithm as well as the reconstruction perform only $O(n)$ \emph{find} calls thus leaving the total running time unchanged. 

Another issue during implementation is that, after contracting a cycle, it is represented by one of its cycle vertices.
The representative is chosen by the DSU among the cycle vertices according to the union-by-size strategy~\cite{Tarjan1975}.
The picked edge incoming to the contracted vertex thus overrides the cycle edge of this representative.
The representative and the edge that was (mistakenly) replaced are saved during the main algorithm and restored in reconstruction just before the actual edge is determined that should be replaced.

\end{document}